\documentclass[conference]{IEEEtran}
\IEEEoverridecommandlockouts
\usepackage{cite}
\usepackage{amsmath,amssymb,amsfonts}
\usepackage{algorithmic}
\usepackage{graphicx}
\graphicspath{ {images/} }
\usepackage{textcomp}
\usepackage{xcolor}
\usepackage{multirow}
\usepackage{algorithmic}
\usepackage{array}
\usepackage{booktabs}
\usepackage{multirow}
\usepackage{enumitem}
\usepackage{listings}
\usepackage{tabularx}
\usepackage{array}
\usepackage{ragged2e}
\def\BibTeX{{\rm B\kern-.05em{\sc i\kern-.025em b}\kern-.08em
    T\kern-.1667em\lower.7ex\hbox{E}\kern-.125emX}}

\newcommand{\toolontology}{\textit{Tool Ontology }}

\newcommand{\inlinedsectionit}[2][6pt]{\vspace{#1}\noindent\textit{#2}.}

\newcommand{\proj}{\textsc{Aegis}}

\begin{document}
\title{\normalsize\normalfont Preprint for 2026 56th Annual IEEE International Conference on Dependable Systems and Networks - Supplemental Volume (DSN-S)\\[1em]
\LARGE \proj: Preventing Cross-Domain Resource Abuse in MCP}

\author{\IEEEauthorblockN{Shriti Priya}
\IEEEauthorblockA{
\textit{IBM Research, USA}\\
shritip@ibm.com}
\and
\IEEEauthorblockN{Teryl Taylor}
\IEEEauthorblockA{
\textit{IBM Research, USA}\\
terylt@ibm.com}
\and
\IEEEauthorblockN{Frederico Araujo}
\IEEEauthorblockA{
\textit{IBM Research, USA}\\
frederico.araujo@ibm.com}
}

\maketitle

\advance\baselineskip-.0pt plus.1pt minus.2pt

\begin{abstract}
The Model Context Protocol (MCP) is an open-source JSON-RPC protocol that standardizes how large language models (LLMs) interact with external systems through programmatic functions known as tools. Attackers or malicious agents can exploit certain modalities of these MCP tools to degrade the overall quality of service of agent-based applications. For example, an agent may request an excessively large search radius or very long videos, overloading backend systems and potentially causing slowdowns or denial-of-service. Each modality---including text, images, video, and location---introduces distinct vectors for resource abuse, complicating the development of consistent mitigation strategies. Moreover, multimodal and cross-domain tools expose diverse request schemas and parameters, making it difficult to define policies that are both generalizable and precise enough to enforce meaningful resource constraints.

In this paper, we present \proj{}, a policy enforcement component that enables administrators to define fine-grained safeguards against resource abuse across heterogeneous MCP tools and modalities. \proj{} leverages the reasoning capabilities of large language models to analyze, categorize, and normalize diverse tool invocations into a unified, policy-friendly representation accessible to security practitioners. Integrated with the Open Policy Agent and the ContextForge AI Gateway, \proj{} detects and mitigates abusive behaviors while preserving the flexibility of MCP-based agent ecosystems.

\end{abstract}

\begin{IEEEkeywords}
Agentic security, Resource abuse, MCP, Policies, LLMs
\end{IEEEkeywords}

\section{Introduction}
\label{sec:1}

The Model Context Protocol (MCP)~\cite{mcp-protocol} defines a standardized JSON-RPC interface through which large language models (LLMs) interact with external systems via programmatic tools. As the number of MCP servers continues to grow---driven by the flexibility and integration advantages they offer---the attack surface of MCP-based systems expands accordingly. One emerging risk is resource exhaustion caused by the parametric exploitation of tools. 

A recent study by EnkryptAI~\cite{enkryptai2026} evaluated 1,000 MCP servers and found that 15\% were vulnerable to resource exhaustion issues, including missing pagination limits, unbounded memory loops, memory leaks, and servers that could be brought down by a single malicious request. The study highlighted \texttt{kubernetes-mcp-server} as an illustrative case, where unconstrained parameters such as \texttt{replicas}, \texttt{timeout}, and \texttt{label\_selector} can trigger excessive resource consumption within a Kubernetes cluster, ultimately leading to service degradation or complete denial of service.

Prior work has explored input validation~\cite{wang2025mcp}, resource quotas~\cite{narajala2026enterprise}, and runtime enforcement mechanisms~\cite{xing2025mcp} to mitigate such issues in MCP servers. Despite these efforts, new MCP servers continue to be deployed rapidly. A widely adopted runtime enforcement approach is the MCP Gateway, which acts as an intermediary layer that validates interactions between agents and MCP servers. These gateways host a diverse set of servers that agents access based on their operational requirements. To strengthen security at this layer, several systems introduce pluggable policy enforcement mechanisms, including OPA~\cite{opa_2024}, rate-limiting, and Cedar-based plugins~\cite{cutler2024cedar}. While these plugins enable policy-driven enforcement, significant challenges arise when such mechanisms are applied across large and heterogeneous MCP ecosystems.

\inlinedsectionit{Diverse Modalities} Modern MCP servers support multiple modalities---including text, image, audio, video, and spatial data such as location. Agents frequently operate across modalities within a single task. For example, the request ``Find restaurants opened from 2020 to now within 50 miles, cap at 100 results, and generate a table showing each restaurant’s name with images of its top three food specialties at 300 dpi,'' which requires the agent to interact with three MCP servers for location queries, text retrieval, and image generation. Each modality introduces distinct vectors for resource abuse: retrieving 100 text records is far less demanding than retrieving or generating 100 videos. These differences complicate the design of consistent policy controls.

\inlinedsectionit{Diverse Parameter Characteristics} The same example involves heterogeneous parameter types, e.g., temporal (2020 to present), volumetric (100 results, three images), and qualitative (50-mile radius, 300 dpi). Each parameter type can trigger excessive resource consumption. Without appropriate bounds, extreme values---such as querying historical data since the year 1800, generating large numbers of high-resolution images, or expanding spatial queries indefinitely---can overload backend services. Effective policies must therefore account for parameter characteristics and their potential for misuse.

\inlinedsectionit{Diverse Operations} MCP tool invocations may involve generation, retrieval, or deletion operations. Retrieving large datasets can strain storage and network resources, while generating high-resolution images or videos can consume substantial compute. Furthermore, generating 100 images is typically far more resource-intensive than retrieving 100 existing images. These operational differences further complicate the definition of consistent resource-control policies.

\inlinedsectionit{Diverse Parameter Semantics} Across hundreds of MCP servers and registries, parameters that represent the same concept are often named differently. For example, location tools may use parameters such as \texttt{radius}, \texttt{initial\_radius}, or \texttt{end\_position}, while text-based servers may limit results using \texttt{limit}, \texttt{maxResults}, or \texttt{count}. This semantic inconsistency significantly complicates policy authoring and enforcement, forcing policy writers to manually track and normalize heterogeneous parameter vocabularies.

\inlinedsectionit{Policy Complexity Explosion} As a result, policy authors must understand not only the modalities supported by each server but also the parameters and resource-abuse vectors associated with each modality across dimensions such as time, quality, and quantity. Performing this analysis for every tool across a large MCP ecosystem is extremely difficult. It requires deep contextual knowledge of tool semantics and system behavior, making policy development complex, time-consuming, and difficult to scale.

To address these challenges, we present \proj, a security system that leverages the reasoning capabilities of large language models to assist in the creation and enforcement of policies that protect against resource abuse across diverse modalities and heterogeneous MCP servers. The main contributions of this paper are:

\begin{itemize}[leftmargin=*] 
    \item A systematic investigation that identifies key parameters contributing to resource exhaustion across MCP servers.
    
    \item A methodology that uses LLMs to extract and normalize resource-related parameters from tool invocations into a unified taxonomy, enabling reusable policy templates and scalable policy enforcement across heterogeneous systems.
    
    \item A threshold estimation approach that provides end-to-end protection against resource abuse attacks in MCP servers.
    
\end{itemize}

\section{Background and Related Work}

The Model Context Protocol (MCP)~\cite{mcp-protocol} is an open protocol that enables AI agents to discover and connect to external tools, data sources, and services through a client–server architecture.
MCP supports multimodal data through typed content blocks (e.g., text, image, audio) and binary resources shared using MIME types and base64-encoded blobs. In production deployments, MCP gateways are often placed in front of multiple MCP servers to provide a unified tool interface. A gateway enables centralized routing, authentication, access control, and observability. Examples include the ContextForge AI Gateway~\cite{mcp_context_forge} and TrueFoundry~\cite{truefoundry_mcp_gateway}.

Recent studies~\cite{guo2025systematic, gaire2025systematization} highlight resource abuse as an emerging security concern in agent-driven systems. Because agents autonomously generate tool calls and parameters, seemingly valid requests may still trigger expensive operations, potentially leading to denial-of-service, increased operational costs, or cascading failures in distributed environments. The OWASP \emph{Agentic AI Threat and Mitigation} report~\cite{owasp_agentic_ai_threats} similarly identifies resource overload as a key risk when agents overuse compute, network, or data resources.

Empirical studies~\cite{zhou2026beyond,lee2026overthinking} show that malicious tool metadata or cyclic registries can significantly inflate token usage and runtime while still producing apparently correct outputs. \emph{LeechHijack}~\cite{zhang2025leechhijack} further demonstrates that MCP tools can parasitize compute budgets without violating explicit permissions. At the ecosystem level, measurement studies such as MCPDiFF~\cite{li2026don} and the large-scale MCP server study~\cite{hasan2025model} reveal that description–implementation mismatches and weak operational practices can conceal expensive or risky behavior.

Benchmark suites including MCPSecBench~\cite{yang2025mcpsecbench}, MCPTox~\cite{wang2025mcptox}, and MCP Security Bench expand evaluation coverage of MCP attack surfaces. Several defenses have also been proposed. Systems such as MCP Guardian~\cite{kumar2025mcp}, MCP-specific penetration testing~\cite{siameh2025context}, MindGuard~\cite{wang2025mindguard}, ETDI~\cite{bhatt2025etdi}, MCP-Guard~\cite{xing2025mcp}, and MCIP~\cite{jing2025mcip} introduce techniques including rate limiting, resource quotas~\cite{narajala2026enterprise}, provenance tracking~\cite{wang2025mcptox}, signed tool definitions~\cite{bhatt2025etdi,jamshidi2026secure}, and loop-aware monitoring. However, the literature still lacks standardized cost semantics for tool schemas and practical trajectory-level verification for real deployments. Other work~\cite{errico2025securing} further shows that rate limiting alone is insufficient when workflows cascade across multiple tools, motivating cost-aware and workflow-aware resource controls.

None of these approaches address the \emph{fine-grained, parameter-specific resource controls} that \proj{} introduces \emph{across heterogeneous and multimodal MCP servers}. Prior work mainly enforces static constraints, such as string length limits or numerical bounds, via resource quotas. In contrast, \proj{} applies dynamic, parameter-aware controls at runtime, enabling policies that incorporate argument semantics and enforce CPU and memory limits per invocation.

\section{Threat Model}

We consider a distributed agentic system in which an agent workflow interacts with an LLM provider and multiple MCP servers deployed across diverse domains. Each MCP server exposes tools that operate across different modalities. We assume that these tools lack robust parameter validation, creating opportunities for misuse by a compromised agent or legitimate user's unintentional excessive request.
%

Our work focuses on the \textit{Resource Overload (T4)} threat identified in the OWASP Agentic AI Threat and Mitigation report~\cite{owasp_agentic_ai_threats}. In this threat model, an attacker or misuse scenario targets the computational, memory, or service capacities of AI systems in order to degrade performance or cause partial system failures. 
We do not consider network-level denial-of-service attacks or vulnerabilities within the MCP protocol itself; our focus is on misuse of tool parameters and resource-intensive invocations. 
The scope excludes high-volume request flooding and instead targets individual requests with excessive or unbounded parameter values that can trigger disproportionate resource consumption.


\section{Approach}
\label{sec:conceptual_overview}

Our approach addresses resource abuse in MCP servers through a \toolontology that formalizes and simplifies tool definitions. The ontology categorizes tools and their parameters, enabling reusable policies and thresholds that prevent resource abuse. This section introduces the ontology and policy model; the next section presents the system architecture.

\subsection{Tool Ontology}

\begin{figure*}
\centering
\includegraphics[width=\linewidth,scale=1]{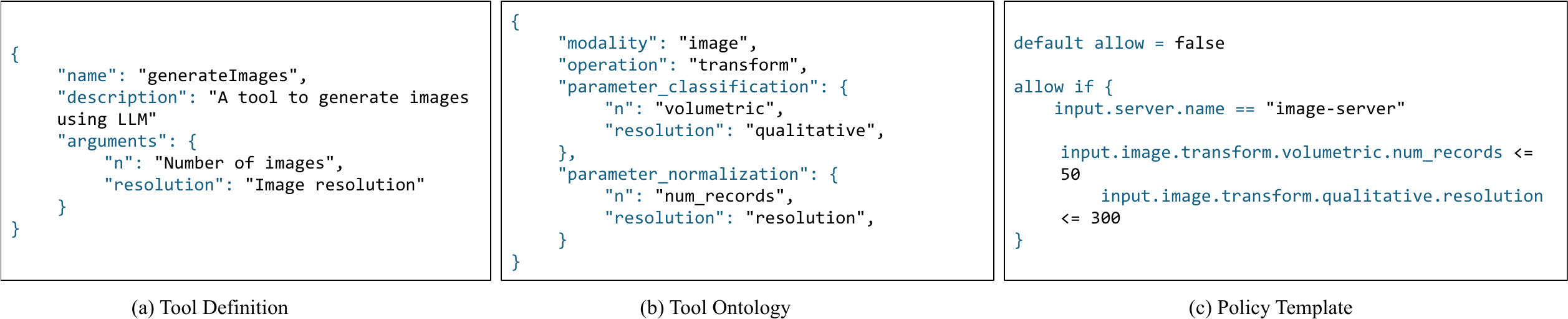}
\caption{From tool definition to policy template}
\label{fig:request_flow}
\end{figure*}

Figure~\ref{fig:request_flow}(a) illustrates an MCP tool definition for a \texttt{generateImages} tool that generates \texttt{n} images at a specified \texttt{resolution}. The figure also shows the corresponding \textit{Tool Ontology}, which extracts the following information from the tool definition.

\inlinedsectionit{Modality}
The primary data type processed by a tool (e.g., image, video, audio, text, location, or human-in-the-loop input). For example, \texttt{generateImages} is classified as an image modality.

\inlinedsectionit{Operation}
The action performed by the tool. We classify tools into three categories: \emph{transform} (create or modify data), \emph{retrieve} (access existing data), and \emph{delete} (remove data). These distinctions capture different resource profiles. For instance, generating 1{,}000 images is significantly more expensive than retrieving 1{,}000 existing records. Similarly, retrieving video data consumes more bandwidth and compute than retrieving text. The \texttt{generateImages} tool is therefore classified as a transform operation.

These distinctions also guide policy measurement units. Text retrieval may use a count-based parameter, whereas video retrieval may require time-based thresholds (e.g., duration in seconds).

\inlinedsectionit{Parameter Characteristics}
Tool parameters are categorized according to their role:
\begin{itemize}[leftmargin=*]
\item \textit{Volumetric}: controls data volume (e.g., \texttt{limit}, \texttt{count})
\item \textit{Timebound}: controls temporal extent (e.g., video duration)
\item \textit{Qualitative}: controls output quality (e.g., image resolution)
\end{itemize}

\noindent In the \texttt{generateImages} example, parameter \texttt{n} is volumetric (number of images), while \texttt{resolution} is qualitative.

\inlinedsectionit{Parameter Normalization}
To support reusable policies, parameters are normalized into a set of common names across tools and servers. We derived this vocabulary by manually analyzing 56 widely used MCP servers mentioned in Section~\ref{sec:dataset}, examining their tool definitions, and identifying parameters associated with resource abuse across modalities. The resulting unified parameter taxonomy is shown in Table~\ref{tab:resource_abuse_parameters}. In Figure~\ref{fig:request_flow}(b), for example, parameter \texttt{n} is normalized to \texttt{num\_records}.


\begin{table}[t]
\scriptsize
\centering
\caption{Resource abuse parameters across modalities}
\label{tab:resource_abuse_parameters}
\newcolumntype{P}[1]{>{\RaggedRight}p{#1}}
\begin{tabular}{lP{6.8cm}}
\toprule
\textbf{Modality} & \textbf{Parameters} \\
\midrule
Image & Aspect Ratio (\texttt{aspect\_ratio}), number of images (\texttt{num\_records}), scaling percent (\texttt{scaling\_percent}), width (\texttt{width}), height (\texttt{height}), channels (\texttt{num\_channels}), resolution (\texttt{resolution}) \\
\midrule
Video/Audio & Start time (\texttt{start\_time}), end time (\texttt{end\_time}), bitrate (\texttt{bitrate}), resolution (\texttt{resolution}), number of videos (\texttt{num\_records}), channels (\texttt{num\_channels}), format (\texttt{format}), sampling rate (\texttt{sampling\_rate}) \\
\midrule
Text & Size, start/end time (\texttt{start\_time}, \texttt{end\_time}), record count (\texttt{num\_records}), filters (\texttt{filter\_array}), repeat interval (\texttt{repeat\_interval}), timeout (\texttt{timeout}), breadth/depth (\texttt{max\_breadth}, \texttt{max\_depth}), retries (\texttt{retries}) \\
\midrule
Location & Radius (\texttt{radius}), start position (\texttt{start\_position}), end position (\texttt{end\_position}), waypoints (\texttt{waypoints}), precision (\texttt{precision}) \\
\midrule
Human & Number of users (\texttt{num\_users}) \\
\bottomrule
\end{tabular}
\end{table}

\subsection{Policy Templates}
\label{subsec:policy_template}

Once parameters are normalized, policies can be expressed using generic templates such as the one shown in Figure~\ref{fig:request_flow}(c). Policies reference parameters using dot notation derived from the ontology layers described above. This abstraction decouples policies from individual tool implementations, reduces policy complexity, and enables reuse across heterogeneous MCP servers. 

\subsection{Policy Thresholds}
\label{subsec:policy_threshold}

Policy templates require carefully chosen thresholds to prevent resource abuse without disrupting normal operation. Threshold selection is challenging because servers handling different modalities (e.g., text, image, or audio) exhibit different baseline CPU and memory usage.

We therefore derive thresholds using an acceptable tool-call error rate as the stability criterion. Parameter values are increased until system performance degrades beyond the specified error tolerance. The highest value that maintains stable operation is selected as the final threshold and incorporated into the policy.

\section{\proj{} System Architecture}

\proj{} generates tool ontologies from tool definitions, constructs policy templates, and estimates policy thresholds. The resulting policies are enforced through an MCP gateway.
\proj{} operates in two phases: \emph{Bootstrapping} and \emph{Runtime}.

\begin{figure}[t]
\centering
\includegraphics[width=\linewidth,scale=1]{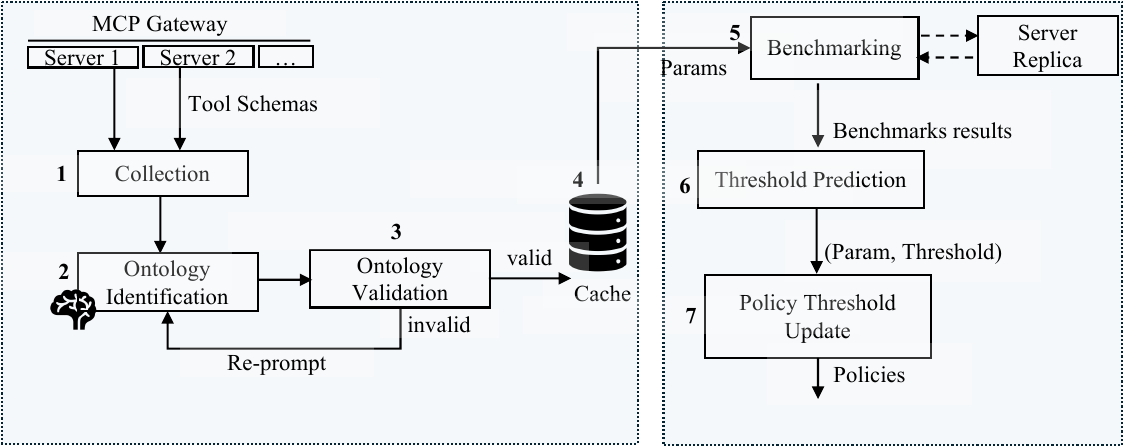}
\caption{\proj{} Bootstrapping}
\label{fig:plugin_bootstrapping}
\vspace{-15pt}
\end{figure}

\subsection{Bootstrapping Phase}

Bootstrapping is an offline process that prepares policies before deployment. Figure~\ref{fig:plugin_bootstrapping} illustrates the bootstrapping phase of \proj{}, which consists of seven steps described below:

\inlinedsectionit{Collection (Step 1)}
At first \proj{} retrieves tool definitions registered with the MCP gateway using the \texttt{list\_tools()} API, which returns tool schemas and metadata.

\begin{figure}[t]
\centering
\includegraphics[width=\linewidth,scale=1]{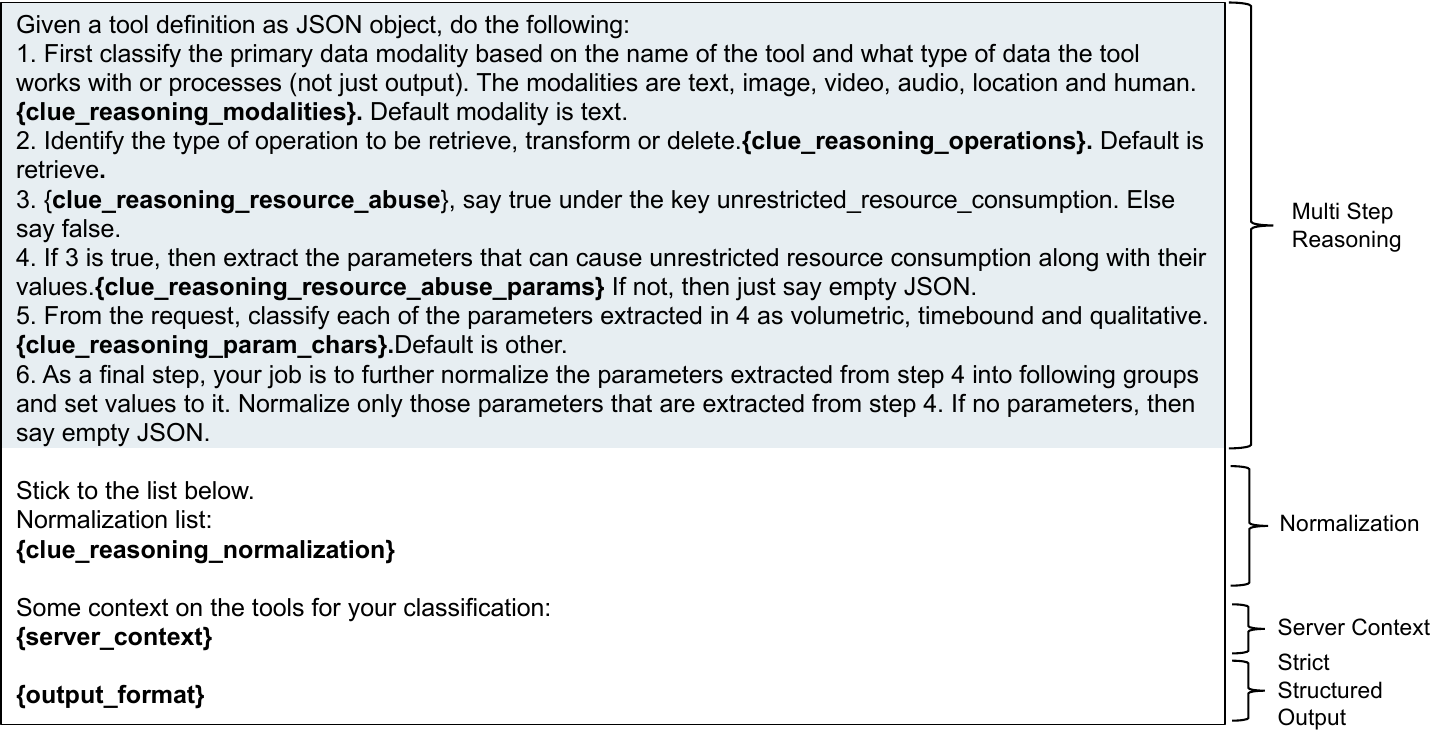}
\caption{System prompt for ontology identification}
\label{fig:prompt_structure}
\vspace{-10pt}
\end{figure}

\inlinedsectionit{Ontology Identification (Step 2)}
Each tool definition is processed by an LLM using a structured system prompt (Figure~\ref{fig:prompt_structure}) to generate a corresponding tool ontology. The prompt guides the model through sequential reasoning using generalized clues and reasoning patterns derived from our analysis of the 937 tool definitions described in Section~\ref{sec:dataset}. The system prompt consists of four components:

\begin{itemize}[leftmargin=*]

\item \textit{Multi-step reasoning}:  
The model performs six sequential steps: (1) classify the tool modality, (2) identify the operation type, (3) determine whether parameters may cause resource abuse, (4) identify resource-related parameters, (5) classify parameter characteristics, and (6) normalize parameters using the predefined taxonomy in Table~\ref{tab:resource_abuse_parameters}.  
Each step is supported by curated clue–reasoning pairs (Figure~\ref{fig:prompt_structure}). For example, the variable \texttt{clue\_reasoning\_modalities} contains patterns such as: \emph{“Requests involving screenshots, image or chart generation (e.g., bar, pie, line, funnel, violin, combo, Sankey), diagrams, social media or other visual content retrieval, publishing, analysis, editing, or transformation of images.”} %
These clues help guide the model toward the intended task and reduce false positives. Similar definitions are provided for all remaining steps.

\item \textit{Normalization}:  
The prompt includes the predefined parameter normalization taxonomy (Table~\ref{tab:resource_abuse_parameters}), along with associated clues and reasoning patterns used to identify and map parameters from tool definitions to the normalized schema.

\item \textit{Server context}:  
Some tools rely on server-specific terminology whose meaning may not be globally obvious. For example, on a Pinterest server, the term \emph{pin} refers specifically to an image. To ensure accurate interpretation, additional contextual information is provided through the \texttt{server\_context} variable (Figure~\ref{fig:prompt_structure}). This context may be manually curated or generated automatically using an LLM or RAG pipeline.

\item \textit{Strict structured output}:  
The LLM outputs the tool ontology (Figure~\ref{fig:request_flow}(b)) as a strict JSON schema. Enforcing a structured format reduces hallucinations, simplifies parsing, and enables automated structural validation.

\end{itemize}




\inlinedsectionit{Ontology Validation (Step 3)}
\proj{} validates the LLM output by ensuring it conforms to the expected JSON structure, verifying that normalized keys match the predefined taxonomy, and confirming that extracted parameters appear in the original tool definition. Invalid outputs trigger re-prompting or are flagged for human review.

\inlinedsectionit{Ontology Cache (Step 4)}
Ontology identification is performed offline to avoid runtime latency. Validated ontologies are cached so that normalized parameters can be applied during policy evaluation. The cache key includes the server name, tool name, and parameters. When new servers are registered, their tools are automatically analyzed and added to the cache.

\inlinedsectionit{Benchmarking (Step 5)}
To estimate policy thresholds, \proj{} benchmarks servers using the identified resource-related parameters. A load generator varies parameter values and concurrent user counts while measuring latency, throughput, error rate, CPU, and memory usage. Benchmarking may run from several hours to multiple days depending on the server and parameter space.

\inlinedsectionit{Threshold Prediction (Step 6)}
Benchmark results are analyzed to determine parameter thresholds. The system selects the highest parameter value that maintains error rates within the acceptable tolerance across all load levels.

\inlinedsectionit{Policy Threshold Update (Step 7)}
The computed thresholds are inserted into the policy templates for each server. Once configured, the policies are ready for deployment in the runtime phase.

\begin{figure}[t]
\centering
\includegraphics[width=\linewidth,scale=1]{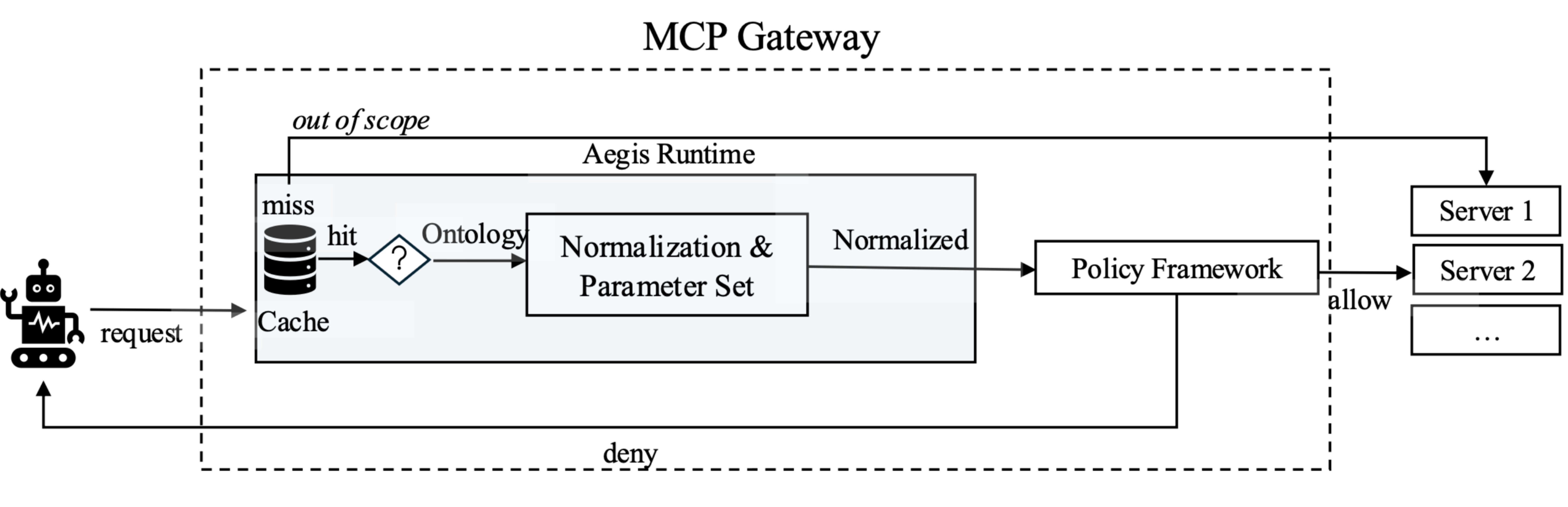}
\caption{\proj{} Runtime}
\label{fig:runtime}
\vspace{-15pt}
\end{figure}

\subsection{Runtime Phase}
Figure~\ref{fig:runtime} illustrates the runtime phase of \proj{}. At runtime, policies are enforced using a framework such as Open Policy Agent (OPA). During a tool invocation, \proj{} checks whether the tool ontology exists in the cache. If present, the request parameters are normalized and evaluated against the policy. Requests within threshold limits are forwarded to the MCP server; otherwise they are rejected.

If the ontology is not cached, the request bypasses normalization and is considered \emph{out of scope} for policy enforcement. 

\section{Implementation}
 The runtime component of \proj{} was implemented using the ContextForge AI Gateway~\cite{mcp_context_forge}. The system leverages the gateway’s plugin framework, which enables modular extension and policy control. To manage policy enforcement, the pre-existing OPA (Open Policy Agent) plugin was adopted as the foundational enforcement layer. 
In this configuration, each tool invocation request is first processed by \proj{} before being forwarded to OPA. 


During the bootstrapping phase \proj{}  collects all the tool definitions of the registered servers in the gateway running ontology identification using the {Claude 4 Sonnet}~\cite{claude_model} LLM. Ontologies are cached in the Redis Database.

\section{Evaluation}

We evaluate \proj{} through the following research questions.

\inlinedsectionit{RQ1}
How accurately can \proj{} identify and categorize tool ontologies from MCP tool definitions?

\inlinedsectionit{RQ2}
How effective is \proj{} at protecting MCP servers against resource abuse attacks?

\subsection{Dataset}\label{sec:dataset}

Since no publicly available labeled dataset exists for this task, we curated a dataset from 56 MCP servers spanning diverse modalities in the OpenTools MCP Registry~\cite{OpenToolsRegistry}. Across these servers, we collected 937 tool definitions. The corresponding tool schemas were crawled from the registry, and a custom parser extracted relevant metadata including tool names, descriptions, parameters, and associated server information.

The dataset covers a broad set of real-world MCP categories, including Cloud Infrastructure (AWS, Azure, Alibaba, Qiniu), Database and Data Management (Pinecone, Meilisearch, YDB, ClickHouse), Productivity and Collaboration (Notion, Sanity, Backlog), Search and Web tools (Tavily, Exa Search, Browserbase, DataForSEO), API and E-commerce platforms (Stripe, Shopify Dev, Chargebee Agentkit), Analytics and Gaming services (OP.GG, Graphlit, PubNub, AntV Chart), and Development tools (E2B, Semgrep, Metricool).

Each of the 937 tools was manually reviewed and assigned a primary modality label from the following categories: text, image, human, video, audio, or location. A tool was labeled according to the primary data modality it processes (e.g., image generation or resizing tools were labeled as image). Tools not associated with image, video, audio, location, or human interaction were labeled as text.

We further annotated each tool by operation type, classifying them as \emph{retrieve}, \emph{transform}, or \emph{delete}. Next, we identified whether a tool contained parameters capable of causing unrestricted resource consumption. A tool was labeled as resource-sensitive if at least one parameter could potentially lead to excessive resource usage.

The extracted parameters were then categorized according to their characteristics as \emph{volumetric}, \emph{qualitative}, or \emph{time-bound}. Finally, all parameter names were normalized using the predefined parameter taxonomy described in Section~\ref{sec:conceptual_overview} Table~\ref{tab:resource_abuse_parameters}.

\inlinedsectionit{Dataset Insights}
Among the 937 tool definitions, 869 belong to the text modality, 38 to image, 14 to location, 10 to video, 5 to audio, and 1 to human.

In terms of operation type, 572 tools were classified as retrieve operations, 308 as transform operations, and 57 as delete operations. Out of all tools, 441 contained parameters capable of causing potential resource abuse.

\subsection{Accuracy (RQ1)}
To evaluate the bootstrapping accuracy of \proj{}, we compared its ontology identification results against the manually labeled ground truth using standard classification metrics including accuracy, precision, recall, and F1-score (Table~\ref{tab:compiled_metrics_simple}).

Experiments were conducted using Claude 4 Sonnet LLM and repeated across three runs with default inference settings with each inference taking approximately 0.52 seconds.
Overall results show more than 84\% accuracy across all ontology identification tasks. The model performed best on modality, operation, parameter characteristics classification and resource consumption detection tasks.  Parameter normalization has lower performance metrics as compared to other tasks due to the complexity of the problem. Misclassifications primarily occurred in cases requiring additional contextual interpretation. For example, distinguishing between storing a binary image file and posting an image on a social media platform requires understanding platform semantics. In systems such as Instagram, the term ``post'' may refer to either text or image content, leading to occasional ambiguity in modality classification.

\begin{table}[t]
\scriptsize
\centering
\caption{Ontology Identification Performance Summary (Mean $\pm$ Std across 3 Runs)}
\label{tab:compiled_metrics_simple}
\begin{tabular}{llc}
\toprule
\textbf{Category} & \textbf{Metric} & \textbf{Mean $\pm$ Std} \\
\midrule
\multirow{4}{*}{Modality} 
& Accuracy & 0.9868 $\pm$ 0.0018 \\
& Precision & 0.9879 $\pm$ 0.0016 \\
& Recall & 0.9868 $\pm$ 0.0018 \\
& F1-Score & 0.9872 $\pm$ 0.0017 \\
\midrule
\multirow{4}{*}{Operation} 
& Accuracy & 0.9740 $\pm$ 0.0013 \\
& Precision & 0.9743 $\pm$ 0.0015 \\
& Recall & 0.9740 $\pm$ 0.0013 \\
& F1-Score & 0.9742 $\pm$ 0.0014 \\
\midrule
\multirow{4}{*}{Resource Consumption} 
& Accuracy & 0.9196 $\pm$ 0.0005 \\
& Precision & 0.9684 $\pm$ 0.0012 \\
& Recall & 0.8571 $\pm$ 0.0000 \\
& F1-Score & 0.9094 $\pm$ 0.0005 \\
\midrule
\multirow{3}{*}{Parameter Characteristics Classification} 
& Exact Match & 1.0000 $\pm$ 0.0000 \\
& Key Match & 1.0000 $\pm$ 0.0000 \\
& Partial Match & 1.0000 $\pm$ 0.0000 \\
\midrule
\multirow{3}{*}{Parameter Normalization} 
& Exact Match & 0.8321 $\pm$ 0.0035 \\
& Key Match & 0.8403 $\pm$ 0.0033 \\
& Partial Match & 0.8342 $\pm$ 0.0035 \\
\bottomrule
\multicolumn{3}{l}{\scriptsize N = 937 samples per run}
\end{tabular}
\end{table}

\subsection{Effectiveness Against Resource Abuse Attacks (RQ2)}

To evaluate the effectiveness of \proj{} in mitigating resource abuse attacks, we conducted controlled experiments measuring policy enforcement under varying load conditions. This required empirically deriving thresholds for abuse-prone parameters and testing whether \proj{} could prevent overload while maintaining normal system operation.

For controlled deployment and measurement, we implemented a custom image-based MCP server called \textit{CIFAR-10 FastMCP Server} using the FastMCP framework. The server exposes the CIFAR-10 dataset~\cite{krizhevsky2009learning}, which contains 60,000 $32\times32$ color images across 10 object classes, as AI-callable tools that allow agents to retrieve and manipulate images. Images are stored as raw RGB bytes in a SQLite database (\texttt{cifar10.db}) configured with WAL (Write-Ahead Logging) mode, a 64~MB page cache, and a 30-second lock timeout to support safe concurrent reads. The server exposes nine tools in total. Our experiments focus on the \texttt{get\_images} tool, where the \texttt{count} parameter controls the number of images returned and can be manipulated to induce resource exhaustion.

\begin{figure}[t]
\centering
\includegraphics[width=\columnwidth]{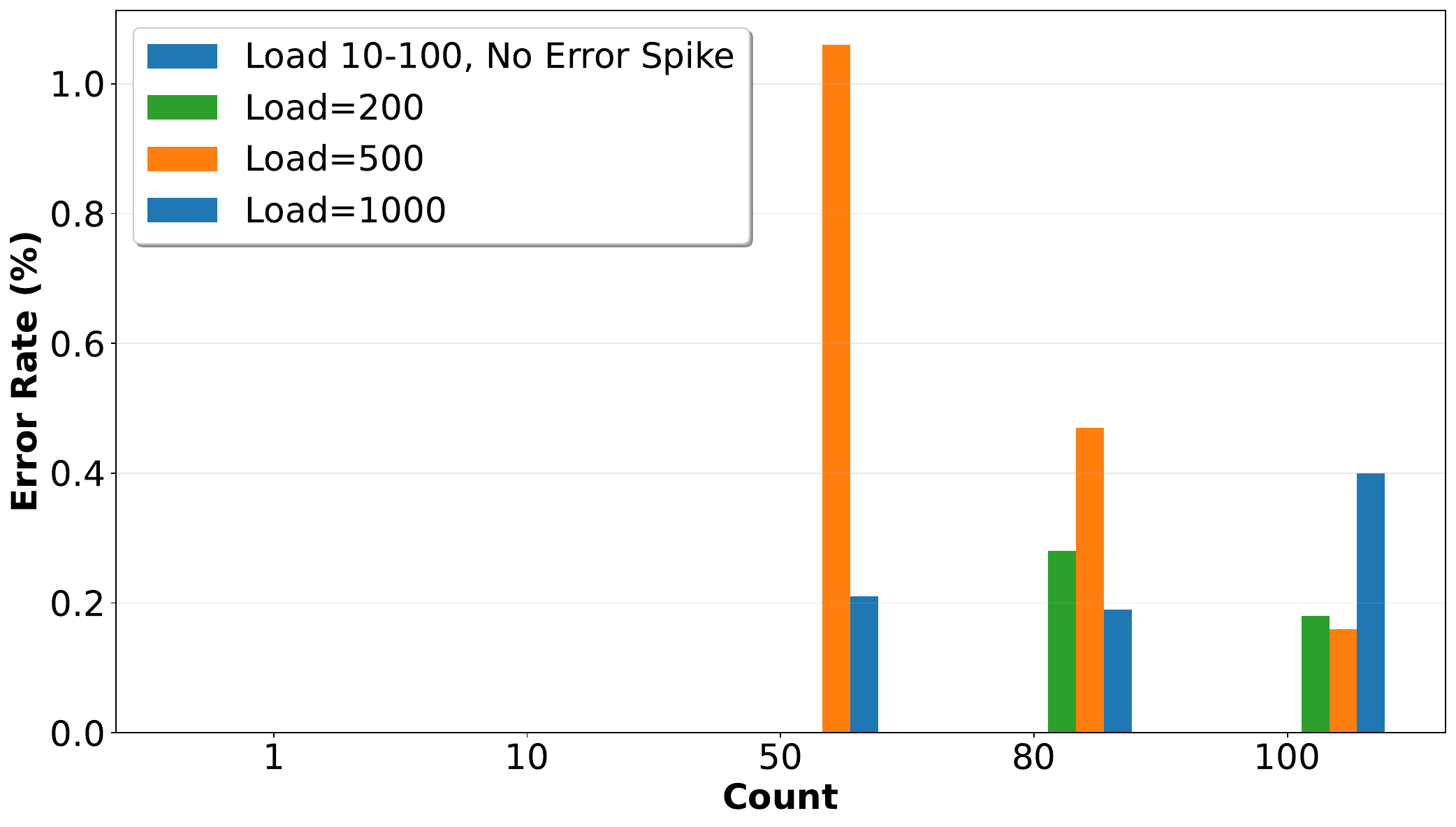}
\caption{The threshold estimation was conducted by benchmarking the CIFAR-10 MCP Server using the \texttt{get\_images} tool, which retrieves images based on specified count parameters. The experiment comprised 118 benchmark runs across varying load levels (10, 25, 50, 100, 200, 500, and 1000) and image count values (1, 10, 50, 80, and 100).}
\label{fig:threshold_estimation}
\vspace{-15pt}
\end{figure}

To estimate policy thresholds, we varied the number of concurrent users (10, 25, 50, 100, 200, 500, 1000) invoking \texttt{get\_images}  and the \texttt{count} parameter (1, 10, 50, 80, 100) as mentioned in Figure~\ref{fig:threshold_estimation}. In total, 118 benchmark runs were executed, with each run taking approximately 8 minutes. The experiments were run on a MacBook Pro (Apple M1 Max chip with 64 GB of RAM).

Experiments used MCP \texttt{ClientSession} with SSE transport.
For each run, we collected the following metrics: latency (mean, min, max, median, p95, p99, standard deviation), throughput (requests/sec), error rate, resource usage (CPU and memory), and total wall-clock execution time. Although multiple metrics were recorded, the primary criterion for selecting parameter thresholds was error rate, as it generalizes well across heterogeneous MCP servers. Following MCP best practices~\cite{mcp_best_practices_2026}, we targeted an error rate below 0.1\%.

\proj{} analyzes the benchmarking results to determine parameter limits that maintain reliable performance under concurrent load. Across the 118 benchmark runs as mentioned in Figure~\ref{fig:threshold_estimation}, \texttt{count=10} consistently maintained error rates below 0.1\% across all load levels, whereas \texttt{count=100} caused significant error spikes. Based on this analysis, \proj{} selects \texttt{count=10} as the policy threshold.

Using this threshold, we then compared server performance with and without \proj{} policy enforcement under a resource abuse scenario. Specifically, we simulated an abuse scenario using \texttt{count=1000} under a 200-concurrent-user load (1 request per user) and measured the impact on key resources in both unprotected and protected configurations as shown in Table \ref{tab:aegis-comparison}. At this high count value, \proj{} effectively maintained server's availability, eliminating errors and improving overall request throughput.

\begin{table}[t]
\scriptsize
\centering
\caption{Detailed Comparison: CIFAR-10 FastMCP Server Performance Without vs. With \proj{}}
\label{tab:aegis-comparison}
\begin{tabular}{lrr}
\toprule
\textbf{Metric} & \textbf{Without \proj{}} & \textbf{With \proj{}} \\
\midrule
Total Load & 200 & 200 \\
Successful & 4 & 200 \\
Failed & 196 & 0 \\
Throughput (req/s) & 1.34 & 30.03 \\
Min Latency (ms) & 1064.84 & 469.79 \\
Median Latency (ms) & 15380.94 & 2366.13 \\
P95 Latency (ms) & 15562.50 & 2863.94 \\
CPU Max (\%) & 24.85 & 29.07 \\
CPU Avg (\%) & 36.40 & 47.84 \\
Memory Max (MB) & 504.77 & 387.72 \\
Memory Avg (MB) & 380.17 & 4.17 \\
Wall Time (s) & 146.27 & 4.17 \\
\bottomrule
\end{tabular}
\vspace{-15pt}
\end{table}

\section {Conclusion}

In this study, we leveraged LLMs to address resource abuse across MCP servers and modalities, achieving over 90\% accuracy in classifying modality, operation, resource usage, and parameters. Parameter normalization remains challenging, highlighting an important area for improvement as LLM capabilities evolve. Our evaluation serves as a proof of concept for ~\proj{} in enabling end-to-end protection, though it is limited in scope and coverage. Future work will extend to real-world MCP services and incorporate cascading tool invocations for more comprehensive protection.
\bibliographystyle{IEEEtran}
\bibliography{references}

@misc{mcp-protocol,
    author = {{Model Context Protocol}},
    title = {{What is the Model Context Protocol ({MCP})?}},
    year=2024,
    url = {https://modelcontextprotocol.io/docs}
}

@misc{enkryptai2026,
  author       = {{Enkrypt AI}},
  title        = {{We Scanned 1,000 {MCP} Servers: 33\% Had Critical Vulnerabilities}},
  year         = {2025},
  url = {https://enkryptai.com/blog/we-scanned-1-000-mcp-servers-33-had-critical-vulnerabilities}
}

@article{wang2025mcp,
  author={Wang, Zhenting and Chang, Qi and Patel, Hemani and Biju, Shashank and Wu, Cheng-En and Liu, Quan and Ding, Aolin and Rezazadeh, Alireza and Shah, Ankit and Bao, Yujia and others},
  title        = {{MCP-Bench: Benchmarking Tool-Using LLM Agents with Complex Real-World Tasks via MCP Servers}},
  year         = {2025},
  url          = {https://doi.org/10.48550/arXiv.2508.20453},
journal={arXiv preprint arXiv:2508.20453}
}

@inproceedings{narajala2026enterprise,
  title={{Enterprise-grade security for the model context protocol ({MCP}): Frameworks and mitigation strategies}},
  author={Narajala, Vineeth Sai and Habler, Idan},
  booktitle={2026 IEEE 5th International Conference on AI in Cybersecurity (ICAIC)},
  pages={1--8},
  year={2026},
  organization={IEEE},
  url={https://ieeexplore.ieee.org/document/11395723}
}

@article{xing2025mcp,
  author={Xing, Wenpeng and Qi, Zhonghao and Qin, Yupeng and Li, Yilin and Chang, Caini and Yu, Jiahui and Lin, Changting and Xie, Zhenzhen and Han, Meng},
  title        = {{MCP-Guard: A Defense Framework for Model Context Protocol Integrity in Large Language Model Applications}},
  year         = {2025},
  url          = {https://doi.org/10.48550/arXiv.2508.10991},
journal={arXiv preprint arXiv:2508.10991}
}

@misc{opa_2024,
  author = {{Open Policy Agent}},
  title = {{Open Policy Agent ({OPA}) Documentation}},
  year={2016},
  url = {https://www.openpolicyagent.org/docs}
}

@article{cutler2024cedar,
  title={{Cedar: A new language for expressive, fast, safe, and analyzable authorization}},
  author={Cutler, Joseph W and Disselkoen, Craig and Eline, Aaron and He, Shaobo and Headley, Kyle and Hicks, Michael and Hietala, Kesha and Ioannidis, Eleftherios and Kastner, John and Mamat, Anwar and others},
  journal={Proceedings of the ACM on Programming Languages},
  volume={8},
  number={OOPSLA1},
  pages={670--697},
  year={2024},
  publisher={ACM New York, NY, USA},
  url={https://doi.org/10.1145/3649835}
}

@misc{mcp_context_forge,
  author = {IBM},
  title = {{ContextForge: Model Context Protocol gateway \& proxy - unify REST, MCP, and A2A with federation, virtual servers, retries, security, and an optional admin UI}},
  year = {2025},
  url = {https://github.com/IBM/mcp-context-forge},
}

@misc{truefoundry_mcp_gateway,
  title = {{MCP Gateway - Secure \& Unified Access to {MCP} Servers}},
  author = {TrueFoundry},
  url = {https://www.truefoundry.com/mcp-gateway},
}

@article{guo2025systematic,
  author={Guo, Yongjian and Liu, Puzhuo and Ma, Wanlun and Deng, Zehang and Zhu, Xiaogang and Di, Peng and Xiao, Xi and Wen, Sheng},
  title        = {{Systematic Analysis of MCP Security}},
  year         = {2025},
  url={https://doi.org/10.48550/arXiv.2508.12538},
journal={arXiv preprint arXiv:2508.12538}
}

@article{gaire2025systematization,
  author={Gaire, Shiva and Gyawali, Srijan and Mishra, Saroj and Niroula, Suman and Thakur, Dilip and Yadav, Umesh},
  title        = {{Systematization of Knowledge: Security and Safety in the Model Context Protocol Ecosystem}},
  year         = {2025},
  url          = {https://doi.org/10.48550/arXiv.2512.08290},
journal={arXiv preprint arXiv:2512.08290}
}

@article{jamshidi2026secure,
  author={Jamshidi, Saeid and Nafi, Kawser Wazed and Dakhel, Arghavan Moradi and Khomh, Foutse and Nikanjam, Amin and Hamdaqa, Mohammad Adnan},
  title        = {{Secure Tool Manifest and Digital Signing Solution for Verifiable MCP
                  and LLM Pipelines}},
  year         = {2026},
  url          = {https://doi.org/10.48550/arXiv.2601.23132},
journal={arXiv preprint arXiv:2601.23132}
}

@article{zhou2026beyond,
  author={Zhou, Kaiyu and Zheng, Yongsen and He, Yicheng and Xue, Meng and Gong, Xueluan and Wang, Yuji and Zhang, Xuanye and Lam, Kwok-Yan},
  title        = {{Beyond Max Tokens: Stealthy Resource Amplification via Tool Calling
                  Chains in LLM Agents}},
  year         = {2026},
  url          = {https://doi.org/10.48550/arXiv.2601.10955},
journal={arXiv preprint arXiv:2601.10955}
}

@article{lee2026overthinking,
  author={Lee, Yohan and Jang, Jisoo and Choi, Seoyeon and Kim, Sangyeop and Choi, Seungtaek},
  title        = {{Overthinking Loops in Agents: A Structural Risk via MCP Tools}},
  year         = {2026},
  url          = {https://doi.org/10.48550/arXiv.2602.14798},
journal={arXiv preprint arXiv:2602.14798}
}

@article{zhang2025leechhijack,
  author={Zhang, Yuanhe and Wang, Weiliu and Zhou, Zhenhong and Wang, Kun and Zhang, Jie and Sun, Li and Liu, Yang and Su, Sen},
  title        = {{LeechHijack: Covert Computational Resource Exploitation in Intelligent
                  Agent Systems}},
  year         = {2025},
  url          = {https://doi.org/10.48550/arXiv.2512.02321},
journal={arXiv preprint arXiv:2512.02321}
}

@article{li2026don,
  author = {Li, Zhihao and Ma, Boyang and Dai, Xuelong and Xu, Minghui and Zhang, Yue and Yan, Biwei and Li, Kun},
  title        = {{Don't believe everything you read: Understanding and Measuring MCP Behavior under Misleading Tool Descriptions}},
  year         = {2026},
  url          = {https://doi.org/10.48550/arXiv.2602.03580},
  journal={arXiv preprint arXiv:2602.03580}
}

@article{hasan2025model,
  author={Hasan, Mohammed Mehedi and Li, Hao and Fallahzadeh, Emad and Rajbahadur, Gopi Krishnan and Adams, Bram and Hassan, Ahmed E},
  title        = {{Model Context Protocol {(MCP)} at First Glance: Studying the Security
                  and Maintainability of {MCP} Servers}},
  year         = {2025},
  url          = {https://doi.org/10.48550/arXiv.2506.13538},
  journal={arXiv preprint arXiv:2506.13538},

}

@article{yang2025mcpsecbench,
  author = {Yang, Yixuan and Gao, Cuifeng and Wu, Daoyuan and Chen, Yufan and Li, Yingjiu and Wang, Shuai},
  title        = {{MCPSecBench: A Systematic Security Benchmark and Playground for
                  Testing Model Context Protocols}},
  year         = {2025},
  url          = {https://doi.org/10.48550/arXiv.2508.13220},
  journal={arXiv preprint arXiv:2508.13220},
}

@inproceedings{wang2025mcptox,
  title={{MCPTox: A Benchmark for Tool Poisoning on Real-World MCP Servers}},
  author={Wang, Zhiqiang and Gao, Yichao and Wang, Yanting and Liu, Suyuan and Sun, Haifeng and Cheng, Haoran and Shi, Guanquan and Du, Haohua and Li, Xiangyang},
  booktitle={Proceedings of the AAAI Conference on Artificial Intelligence},
  volume={40},
  number={42},
  pages={35811--35819},
  year={2026},
  url={https://doi.org/10.1609/aaai.v40i42.40895}
}

@article{errico2025securing,
  title={{Securing the Model Context Protocol (MCP): Risks, Controls, and Governance}},
  author={Errico, Herman and Ngiam, Jiquan and Sojan, Shanita},
  journal={arXiv preprint arXiv:2511.20920},
  year={2025},
  url = {https://doi.org/10.48550/arXiv.2511.20920}
}

@inproceedings{jing2025mcip,
  title={{MCIP: Protecting MCP safety via model contextual integrity protocol}},
  author={Jing, Huihao and Li, Haoran and Hu, Wenbin and Hu, Qi and Heli, Xu and Chu, Tianshu and Hu, Peizhao and Song, Yangqiu},
  booktitle={Proceedings of the 2025 Conference on Empirical Methods in Natural Language Processing},
  pages={1177--1194},
  year={2025},
  url={https://aclanthology.org/2025.emnlp-main.62/}
}

@article{siameh2025context,
  title={{Context injection vulnerabilities and resource exploitation attacks in model context protocol}},
  author={Siameh, Theophilus and Addobea, Abigail Akosua and Liu, Chun-Hung},
  journal={Authorea Preprints},
  year={2025},
  publisher={Authorea},
  url={https://doi.org/10.36227/techrxiv.175321790.02502754/v1}
}

@article{kumar2025mcp,
  title={{MCP Guardian: A security-first layer for safeguarding mcp-based ai system}},
  author={Kumar, Sonu and Girdhar, Anubhav and Patil, Ritesh and Tripathi, Divyansh},
  journal={arXiv preprint arXiv:2504.12757},
  year={2025},
  url = {https://doi.org/10.48550/arXiv.2504.12757}
}

@inproceedings{bhatt2025etdi,
  title={{ETDI: Mitigating tool squatting and rug pull attacks in model context protocol (MCP) by using oauth-enhanced tool definitions and policy-based access control}},
  author={Bhatt, Manish and Narajala, Vineeth Sai and Habler, Idan},
  booktitle={2025 Cyber Awareness and Research Symposium (CARS)},
  pages={1--6},
  year={2025},
  organization={IEEE},
  url={https://ieeexplore.ieee.org/document/11337310}
}

@misc{owasp_agentic_ai_threats,
  author = {{OWASP Foundation: OWASP GenAI Security Project}},
  title = {{Agentic {AI} Threats and Mitigations}},
  year = {2025},
  url = {https://genai.owasp.org/resource/agentic-ai-threats-and-mitigations/},
}

@misc{OpenToolsRegistry,
author = {{OpenTools}},
title = {OpenTools MCP server registry},
year = {2025},
url = {https://opentools.com/registry},
}

@misc{mcp_best_practices_2026,
  author = {{Model Context Protocol}},
  title = {{MCP Best Practices: Architecture \& Implementation Guide}},
  year = {2024},
  url = {https://modelcontextprotocol.info/docs/best-practices/},
}

@misc{claude_model,
    author = {{Anthropic}},
    title = {{Claude Sonnet Models Announcements}},
    year={2025},
    url = {https://www.anthropic.com/claude/sonnet}
}

@misc{krizhevsky2009learning,
    author       = {Krizhevsky, Alex and Hinton, Geoffrey},
    institution  = {University of Toronto},
    title = {{The CIFAR-10 dataset}},
    year={2009},
    url ={https://www.cs.toronto.edu/~kriz/cifar.html}
}

@article{wang2025mindguard,
  title={{MindGuard: Tracking, detecting, and attributing mcp tool poisoning attack via decision dependence graph}},
  author={Wang, Zhiqiang and Zhang, Junyang and Shi, Guanquan and Cheng, HaoRan and Yao, Yunhao and Guo, Kaiwen and Du, Haohua and Li, Xiang-Yang},
  journal={arXiv preprint arXiv:2508.20412},
  year={2025},
  url= {https://doi.org/10.48550/arXiv.2508.20412}

}
\end{document}